\documentstyle[12pt]{article}
\begin{document}
\begin{center}
{\large On Superconductors,Superfluids and Torsion Vortices}
\end{center}
\vspace{2cm}
\begin{center}
by L.C.Garcia de Andrade\footnote{Grupo de Relatividade e Teoria de Campo - 
Instituto de F\'{\i}sica - UERJ, Rio de janeiro, R.J 20550, Brasil} 
\end{center}
\begin{abstract}
\hspace{0.6cm}The Meissner effect for superconductors in spacetimes with torsion is revisited.Two new physical interpretations are presented.The first considers the Landau-Ginzburg theory yields a new symmetry-breaking vacuum depending on torsion.In the second interpretation a gravitational Meissner torsional effect where when the Higgs field vanishes, torsion and electromagnetic fields need not vanish and outside the Abrikosov tubes a torsion vector analogous to the Maxwell potential is obtained.The analogy is even stronger if we think that in this case the torsion vector has to be derivable from a torsion potential.Another solution of Landau-Ginzburg equation is shown to lead naturally to the geometrization of the electromagnetism in terms of the torsion field.As an application we show that Landau-Ginzburg equations extended to Cartan spaces allow us to determine torsion in the case when the order parameter is frozen to a minimum.It is shown that the presence of torsion in solid state physics changes the ground state of the superfluid phases.
\end{abstract}
\vspace{2cm}
\begin{center}
\underline{PACS:} 11.15 EX: Spontaneous breaking of gauge symmetries
\end{center}
\newpage
\hspace{0.6cm}
Ealier D'Auria and Regge \cite{1} have considered gravity theories with torsion and gravitationally asymptotically flat instantons.The vanishing of the gap inside a flux tube had a perfect analogy in the vanishing of the vierbein inside the gravitational instanton.Connection between torsion vortices and non-trivial topological configurations have also been pointed out in \cite{2}.In the following an argument considering the analogy between the Meissner effect and vanishing torsion in gravity is revisited.We shown that two distinct physical interpretations may be given to the Landau-Ginzburg theory with torsion.In the first torsion shifts the symmetry breaking vacuum by a torsion dependent value.In the second the Landau-Ginzburg equation is solved and a torsion potential is obtained outside the Abrikosov tube in analogy with the Maxwell field derivable from a scalar electromagnetic potential in the usual Meissner effect.Finally an application is considered by computing the photon mass in the case of the magnetic field of a neutron star.Let us now considered the Landau-Ginzburg equation for superconductors
\begin{equation}
D_{\mu}D^{\mu}{\phi}=g{\phi}(|\phi|^{2}-{\lambda}^{2})
\label{1}
\end{equation}
where ${\phi}$ is the gap field and g is a coupling constant.
In the non-gravitational Meissner effect near the symmetry-breaking 
vacuum the magnitude of ${\phi}$ is constant
\begin{equation}
|{\phi}|={\lambda}
\label{2}
\end{equation}
By considering the extended covariant derivative to include 
torsion
\begin{equation}
D_{\mu}={\partial}_{\mu}-ieA_{\mu}-ifQ_{\mu}
\label{3}
\end{equation}
where $ Q_{\mu} $ are the components of the torsion vector.
Substitution of the (\ref{3}) into (\ref{1}) yields
\begin{equation}
{D^{em}}_{\mu}{D^{em}}^{\mu}{\phi}=g{\phi}{|\phi|^{2}-({\lambda}^{2}+\frac{f}{g}Q^{2}){\phi}}
\label{4}
\end{equation}
Here $D^{em}$ is the electromagnetic part of the covariant 
derivative given in (\ref{3}).We also have made use of the 
following gauges $D^{em}_{\mu}Q^{\mu}=0$ and $Q^{\mu}D^{em}_{\mu}{\phi}=0$.
Thus it is easy to see from this equation that the 
symmetry-breaking vacuum is shift by a torsion energy term 
$Q^{2}=Q_{\mu}Q^{\mu}$.Situations like that may certainly appear 
in other problems in field theory like domain walls with torsion 
\cite{3,4} and problems in Superfluids \cite{5}.Let us now solve 
the equation (\ref{4}).Suppose that we work in the static case 
where we are from the wall and impurities and the general solution drifts into an asymptotic regime in which 
it is covariantly constant
\begin{equation}
D_{\mu}{\phi}={\partial_{\mu}-ieA_{\mu}-ifQ_{\mu}}{\phi}=0
\label{5}
\end{equation}
Differentiating once again we have
\begin{equation}
{D_{\mu}D_{\nu}-D_{\nu}D_{\mu}}{\phi}=ieF_{\mu \nu}{\phi}
-if(\partial_{\mu}Q_{\nu}-\partial_{\nu}Q_{\mu}){\phi}=0
\label{6}
\end{equation}
Thus the vanishing of the Maxwell tensor is not necessarily followed 
by the vanishing of the torsion.Note that equation (\ref{6}) has two possible solutions.In the first case assuming that the space is free of electromagnetic fields or $F_{\mu\nu}=0$ the vanishing of the second term implies a vortice term given by
\begin{equation}
\partial_{\mu}Q_{\nu}-\partial_{\nu}Q_{\mu}=0
\label{7}
\end{equation}
nevertheless in places of the superconductor where the Higgs field
${\phi}$ vanishes,torsion and electromagnetic fields do not vanish 
as happens on the inside of the Abrikosov tubes.Outside the tube 
equation (\ref{7}) is obeyed and torsion is derived from a torsion 
potential in analogy to the electromagnetic potential
\begin{equation}
A_{\mu}=-\frac{i}{e}\partial_{\mu}\theta
\label{8}
\end{equation}
where the phase ${\theta}$ appears in the solution as 
\begin{equation}
\phi=\lambda e^{i\theta}
\label{9}
\end{equation}
By analogy torsion is given by
\begin{equation}
Q_{\mu}=\frac{i}{f}{\partial}_{\mu}{\alpha}
\label{10}
\end{equation}
where ${\alpha}$ is the torsion phase.To perform the above 
computations we have considered that second order terms in the 
coupling of torsion and the electromagnetic potential may be 
dropped.The second solution assumes that the space is not free of electromagnetic fields and equation (\ref{6}) implies that the electromagnetic fields can be written in terms of torsion as
\begin{equation}
F_{\mu\nu}=\frac{f}{e}(\partial_{\mu}Q_{\nu}-\partial_{\nu}Q_{\mu})
\label{11}
\end{equation}
This implies that the electromagnetic vector potential is proportional to the torsion vector.Similar results have been obtained by R.Hammond \cite{6} in the more general case of the non-Riemannian geometry with curvature and torsion.Let us now consider a non-linear electrodynamics in 
spaces with torsion
\begin{equation}
L\cong\sqrt{-g}[R(\Gamma)(1+A^{2})-\frac{1}{4}F_{ij}F^{ij} + 
J^{i}A_{i}]
\label{12}
\end{equation}
The interaction between the vector potencial and background 
torsion was found to break the gauge invariance leading to a 
Proca type mass term of the form :
\begin{equation}
{m^{2}}_{\gamma}\cong\lambda R(\Gamma)A^{2}
\label{13}
\end{equation}
This term gives the photon a small rest mass.As an application let us consider the Landau-Ginzburg equations \cite{7}
\begin{equation}
[-({\nabla}-ieA-iQ)^{2}+(\frac{T}{T_{c}}-1)+g|{\phi}|^{2}]{\phi}=0
\label{14}
\end{equation}
and
\begin{equation}
{\nabla}X{\nabla}XA =ej
\label{15}
\end{equation}
where $j$ is the three dimensional current and $A$ is the three dimensional electromagnetic vector potential.Here we assume that the presence of torsion does not break the current conservation and in principle is not responsible for any creation of a massive photon.The first Landau-Ginzburg equation can be written in terms of an order parameter ${\rho}$ obtained from the definition ${\phi}={\rho}(x)e^{i{\gamma}(x)}$ as
\begin{equation}
[-({\nabla}-ieA-iQ)^{2}+(\frac{T}{T_{c}}-1)+g|{\rho}|^{2}]{\rho}=0
\label{16}
\end{equation}
and
\begin{equation}
{\nabla}X{\nabla}XA =-e^{2}{\rho}^{2}A(x)
\label{17}
\end{equation}
Separating the real and imaginary parts in the first equation we obtain an equation for ${\rho}$ 
\begin{equation}
[-{\nabla}^{2}+e^{2}A^{2}+f^{2}Q^{2}+(\frac{T}{T_{c}}-1)+g|{\rho}|^{2}]{\rho}=0
\label{18}
\end{equation}
and another for $A$
\begin{equation}
({\partial}_{i}A_{i}){\rho}+2(A_{i}{\partial}_{i}){\rho}+({\partial}_{i}Q_{i}){\rho}+2(Q_{i}{\partial}_{i}){\rho}=0
\label{19}
\end{equation}
This equation is simply solved with some assumptions.It is important to know that equation (\ref{17}) has to be expressed in fact as $({\nabla}-ieA-iQ)X({\nabla}-ieA-iQ)XA=-e^{2}{\rho}^{2}A(x)$ and expansion of this expression considering the torsion as the gradient of a scalar field implies a constraint on the torsion vector S as $SX{\nabla}XA=SXB=0$ where B is the magnetic field.This constraint simply says that the magnetic field has the same direction as the torsion field this situation is interesting when torsion is proportional to spin as in theories of gravity as Einstein-Cartan.This constraint is also fundamental for the charge conservation which implies as usual
\begin{equation}
{\nabla}.j=-e{\nabla}.({\rho}^{2}A)=0
\label{20}
\end{equation}
These equations are easily to solve in the transverse gauge ${\partial}_{i}A_{i}=0$ and in the case where the order parameter
${\rho}$ does not fluctuate or ${\partial}_{i}{\rho}=0$.In this case the above equations reduce to
\begin{equation}
({\partial}_{i}Q_{i}){\rho}=0
\label{21} 
\end{equation}
which is an equation for torsion, and 
\begin{equation}
[e^{2}A^{2}+f^{2}Q^{2}+(\frac{T}{T_{c}}-1)+g|{\rho}|^{2}]{\rho}=0
\label{22}
\end{equation}
The equation (\ref{22}) can be easily solved if we consider that torsion is given by a gradient of a torsion potential ${\psi}$ as $Q={\nabla}{\psi}$.In this case the torsion equation reduces to a simple Laplace equation which in spherical coordinates give a very simple solution like $Q_{r}=\frac{1}{{r}^{2}}$ this torsion implies that the torsion flux is constant.To check the influence of torsion on the order parameter ${\rho}$
let us consider (\ref{22}) in the case the critical temperature is given by
$ T<<<T_{c} $ in this case for ${\rho}$ different from zero we obtain the following approximate value of the order parameter
\begin{equation}
|{\rho}|=[1-\frac{1}{2}(e^{2}A^{2}+f^{2}Q^{2})]
\label{23}
\end{equation}
By considering the electromagnetic vector potential as $|A|=\frac{1}{r}$ and the torsion as calculated above equation (\ref{23}) yields 
\begin{equation}
|{\rho}|=[1-\frac{1}{2}(\frac{e^{2}}{{r}^{2}}+\frac{f^{2}}{r^{4}})]
\label{24}
\end{equation}
As was expected the influence of torsion outside the superconductor is much weaker than the magnetic one.It is also fundamental to see that with the above hyphotesis the current conservation is kept.As a final step we shall demonstrate that the introduction of torsion geometry into condensed matter physics yields a new minima for the potential if we redefine it in terms of the relation between torsion and spin as in Einstein-Cartan gravity.To accomplish such a task we might consider the energy in the superfluid phase, for $T_{c}>T$ with the introduction of torsion in the covariant derivative as above
\begin{equation}
E={\int}dx^{3}(\frac{1}{2}|{\nabla}{\phi}|^{2}+(\frac{{m}^{2}}{2}+f^{2}S^{2})|{\phi}|^{2}+\frac{g}{4}|{\phi}|^{4})
\label{25}
\end{equation}
For $(\frac{{m}^{2}}{2}+f^{2}S^{2})<0$ ,the ground state is given by
\begin{equation}
{\phi}_{0}=(-\frac{{m}^{2}+2f^{2}S^{2}}{g})^{\frac{1}{2}}e^{i{\gamma}_{0}}
\label{26}
\end{equation}
where ${\gamma}_{0}$ is an arbitrary phase angle.Thus we may conclude that the presence of torsion in solid state physics also changes the ground state of the superfluid phases.A detailed investigation of the Landau-Ginzburg equation in Cartan spaces can appear elsewhere.
\section*{Acknowledgements}
I am very grateful to Prof.Kleinert for helpful discussions of the subject of this paper.I also thank Prof.G.Volovik for sending me his papers and for his interest in my work and to the Universidade do estado do Rio de Janeiro
(UERJ) for financial support.

\end{document}